\documentclass[runningheads]{llncs}
\usepackage[T1]{fontenc}
\usepackage{graphicx}
\usepackage[acronym, shortcuts]{glossaries-extra}
\usepackage{graphicx}
\usepackage{multirow}
\usepackage{tikz}
\usepackage[moderate, tracking=normal, leading=normal]{savetrees}
\usepackage{subcaption}
\usepackage{soul}
\usepackage{xcolor}
\usepackage{amsmath}
\usepackage{amssymb}
\usepackage[breaklinks, colorlinks, linkcolor=purple, anchorcolor=purple, citecolor=purple, urlcolor=purple]{hyperref}
\usepackage{booktabs}

\begin{document}
\title{An Attack Method for Medical Insurance Claim Fraud Detection based on Generative Adversarial Network}
%
%
\author{Yining Pang\inst{1}\and
Chenghan Li*\inst{2} }
\authorrunning{Y.Pang et al.}
%
\institute{Department of Mathematics, Hong Kong Baptist University, Kowloon Tong, Hong Kong \and
Tsinghua Shenzhen International Graduate School , Tsinghua University, Shenzhen 518055, China
\email{lich24@mails.tsinghua.edu.cn}\\}
\maketitle              
\begin{abstract}
Insurance fraud detection represents a pivotal advancement in modern insurance service, providing intelligent and digitalized monitoring to enhance management and prevent fraud. It is crucial for ensuring the security and efficiency of insurance systems. Although AI and machine learning algorithms have demonstrated strong performance in detecting fraudulent claims, the absence of standardized defense mechanisms renders current systems vulnerable to emerging adversarial threats. In this paper, we propose a GAN-based approach to conduct adversarial attacks on fraud detection systems. Our results indicate that an attacker, without knowledge of the training data or internal model details, can generate fraudulent cases that are classified as legitimate with a 99\% attack success rate (ASR). By subtly modifying real insurance records and claims, adversaries can significantly increase the fraud risk, potentially bypassing compromised detection systems. These findings underscore the urgent need to enhance the robustness of insurance fraud detection models against adversarial manipulation, thereby ensuring the stability and reliability of different insurance systems.

\end{abstract}

\section{Introduction}
\begin{figure}[h]
\centering
\includegraphics[width=0.9\textwidth]{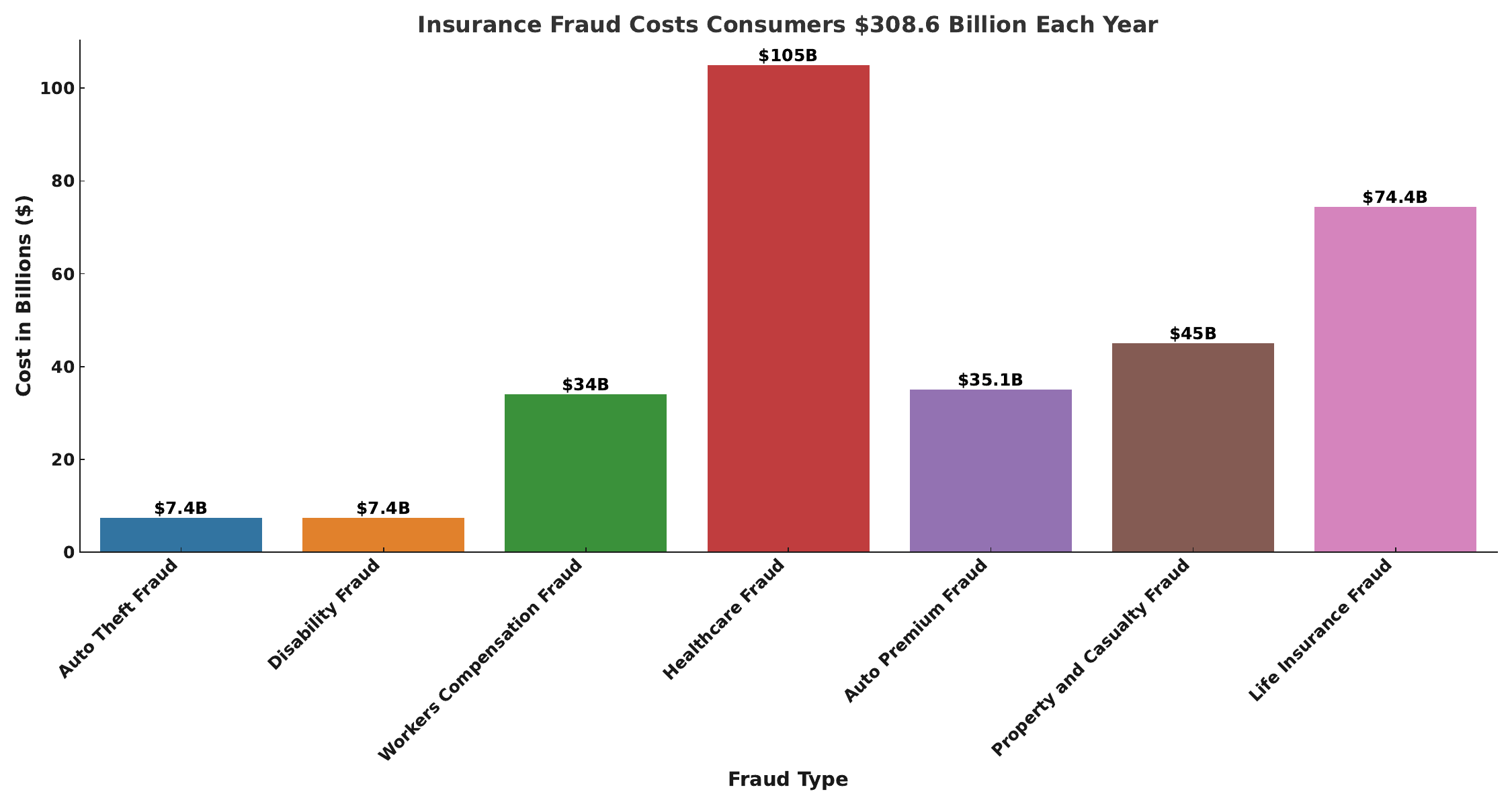} 
\caption{Insurance Fraud Statistics\cite{al2024utilizing}}
\label{fig:cost}
\end{figure}

Insurance fraud poses a significant threat to the insurance industry, particularly in the realm of medical insurance. Medical insurance fraud poses a significant threat to healthcare insurance systems and has increasingly become a focus of public concern. In 2017, healthcare spending in the United States reached approximately \$3.5 trillion \cite{ahn2025implementation}, with over 20\%—around \$720 billion—allocated to medical insurance \cite{jazowski2024comparing}. These substantial expenditures have created opportunities for exploitation by fraudulent individuals and organizations(Figure \ref{fig:cost}). It is estimated that 3\% to 10\% of medical insurance funds—equivalent to \$21 billion to \$71 billion—are lost due to fraudulent activities \cite{bauder2017medicare}. Such fraud not only increases the operational costs of the healthcare system but also imposes additional financial burdens on consumers \cite{pandey2024residual}\cite{khalil2025holistic}\cite{singh2024metaheuristic}.

In the field of insurance fraud detection, traditional machine learning (ML) models have been widely applied, primarily using structured claims data to identify anomalies and fraudulent behaviors \cite{johnson2019medicare,herland2018big}. However, with technological advancements, Agrawal \cite{agarwal2023intelligent} recently proposed an innovative hybrid deep learning (DL) framework that integrates Convolutional Neural Networks (CNNs), Transformer architectures, and XGBoost to enhance detection accuracy and robustness. This framework not only incorporates domain-specific features, such as provider-patient interaction graphs and temporal patterns, but also enhances model interpretability through Shapley Additive exPlanations (SHAP) technology. Evaluated on two datasets---the Medicare provider fraud dataset and the healthcare provider dataset---the model achieved F1 scores of 0.92 and 0.94, respectively, as well as AUC-ROC values exceeding 0.96. Meanwhile, Kasarani \cite{kasaraneni2024machine} conducted an in-depth exploration of various ML methods used in different insurance claim fraud detection. The study categorized ML techniques into supervised learning, unsupervised learning, and anomaly detection, emphasizing the significance of popular models such as Logistic Regression (LR), Random Forests (RFs), and Gradient Boosting in supervised tasks. Additionally, clustering algorithms and Isolation Forests in unsupervised learning were examined. Kasarani highlighted the critical role of feature engineering in enhancing model performance and demonstrated the practical application of ML-based fraud detection systems in various insurances through case studies. The study also addressed challenges such as algorithmic bias, evolving fraud strategies, and data quality issues, proposing strategies to enhance model adaptability and interpretability through explainable AI technologies and domain-specific knowledge. This comprehensive analysis shows that ML techniques can effectively combat insurance fraud by identifying complex patterns and anomalies in claims data, thereby improving detection accuracy and efficiency. On the other hand, Yoo et al. \cite{yoo2023medicare} conducted a comparative study on the application of traditional ML models and Graph Neural Networks (GNNs) in Medicare fraud detection. The study focused on improving fraud detection effectiveness by capturing relationships among providers, beneficiaries, and physicians.This indicates that research on different insurance mechanisms can also enhance insurance fraud detection\cite{cosma2024redefining}\cite{alolayyan2025mediating}.

Adversarial machine learning (AML) \cite{kumar2020adversarial,kurakin2016adversarial,finlayson2019adversarial} is an emerging field that addresses the vulnerabilities of traditional machine learning (ML) models to adversarial manipulation. The concept of adversarial examples—subtle, malicious modifications to input data that can mislead ML models—was first introduced by Chakraborty et al. \cite{chakraborty2021survey}. Since then, several studies have investigated the application of AML in fraud detection scenarios \cite{bortsova2021adversarial,albattah2023detection}. In the context of insurance, adversarial attacks pose a significant risk, as fraudsters can intentionally manipulate claims data to evade detection mechanisms. Research by Chowdhury et al. \cite{chowdhury2024advancing} demonstrated that such attacks can significantly degrade the performance of conventional fraud detection models. To mitigate this issue, adversarial training has been proposed as a defense strategy. Zheng et al. \cite{zheng2020efficient} introduced a method that augments the training dataset with adversarial examples to enhance model robustness and improve the detection of fraudulent claims under attack. Despite these advancements, the application of AML techniques in insurance remains relatively underexplored, highlighting the need for further research to develop resilient and adaptive fraud detection systems.

Through the preceding discussion, it is clear that the security of machine learning methods for fraud detection in various insurance institutions remains underexplored. To address the aforementioned issues, this paper proposes a novel method for attacking insurance fraud detection based on Generative Adversarial Networks (GANs). To the best of our knowledge, this is the first study to propose a GAN-based adversarial attack against insurance fraud, requiring minimal access to real data and models while achieving extremely high success rates. Initially, we evaluated the success rates of various machine learning methods for insurance fraud detection. Subsequently, we trained a GAN to generate attack samples from random data. Furthermore, we assessed the vulnerability of machine learning models to the proposed method under both white-box (i.e., access to models and data) and gray-box (i.e., access to model outputs) scenarios, demonstrating susceptibility even without access to real data or model details. The resulting injected data poses significant risks. The contributions of this paper can be summarized as follows:
\begin{itemize}
    \item We introduce a GAN-based attack method targeting insurance fraud detection, which, to our knowledge, marks the first application of GANs in this specific context.
    \item We develop a generator reinforcement learning training method guided by a surrogate model.
    \item We apply the proposed method to various machine learning models used for insurance fraud detection, and the results validate the effectiveness and robustness of our approach.
\end{itemize}

The remainder of this paper is structured as follows: Section 2 outlines the proposed method and elaborates on the details of the GAN. Section 3 provides an analysis of the dataset and presents the results of the proposed method. Section 4 concludes the paper.

\section{Methods}
In this section, we elaborate on the details of the GAN-based method proposed in this paper. In Section 2.1, we describe the principles and details of GANs. In Section 2.2, we present a surrogate training method based on reinforcement learning. In Section 2.3, we discuss the principles of machine learning models for insurance fraud detection.
\begin{figure}[h]
\centering
\includegraphics[width=0.9\textwidth]{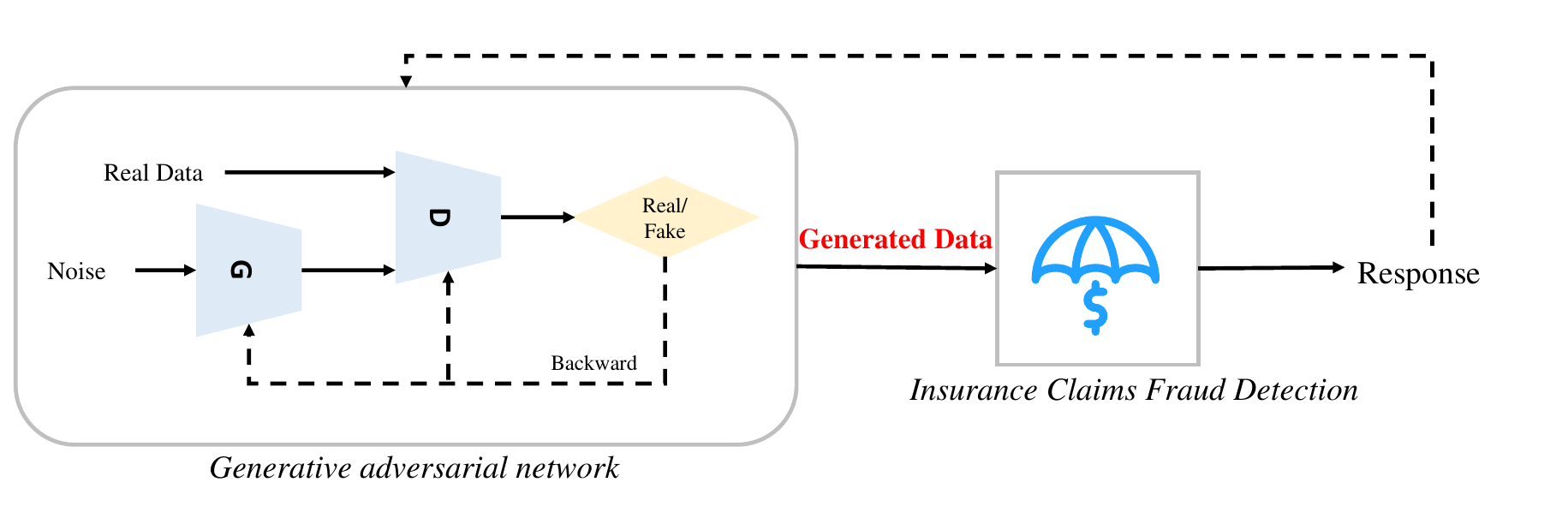} 
\caption{The Framework of GAN Attack Method}
\label{fig:method}
\end{figure}

\subsection{GAN}
Generative Adversarial Networks (GANs)  are a class of deep learning models that have been widely used for unsupervised learning tasks. They consist of two neural networks: a generator and a discriminator, which are trained simultaneously through a zero-sum game framework.The generator $G$ aims to produce synthetic data that is indistinguishable from real data. It takes a random noise vector $z$ as input and generates an output $G(z)$. The discriminator $D$, on the other hand, tries to distinguish between real data $x$ and the synthetic data generated by $G$. The objective of the discriminator is to maximize the probability of correctly classifying both real and generated data.The training process can be formulated as a minimax game with the following objective function:
\begin{equation}
\min_{G} \max_{D} V(D, G) = \mathbb{E}_{x \sim p_{\text{data}}(x)}[\log D(x)] + \mathbb{E}_{z \sim p_z(z)}[\log (1 - D(G(z)))]
\end{equation}
Where $p_{\text{data}}(x)$ is the distribution of real data, and $p_z(z)$ is the distribution of the noise vector.During training, the generator $G$ is updated to minimize the objective function, while the discriminator $D$ is updated to maximize it. This adversarial training process continues until the generator produces samples that are sufficiently realistic to fool the discriminator.

\subsection{RL Training}
Insurance fraud detection can be described as a classification problem. In classification-based adversarial settings, the goal is to generate inputs that cause a target classifier to make incorrect predictions. This task is particularly challenging when access to the target model is restricted, such as in black-box scenarios where model internals (e.g., gradients, architecture) are unavailable. In such settings, surrogate models and reinforcement learning offer a viable path for training an adversarial generator. The adversarial attack is formulated as a generator optimization problem, where inputs that maximize the classification error of a target or surrogate model $S$ are learned to be synthesized by a generator $G$. To handle the black-box nature of the task, a surrogate-guided reinforcement learning framework is employed, where the reward signal is derived from the surrogate's classification output.

The generator receives as input a randomly initialized latent variable $\mathbf{z} \in \mathbb{R}^{B \times T \times F}$, which is transformed into a synthetic sample $\tilde{\mathbf{x}} = G(\mathbf{z})$. This process is repeated over an episodic horizon, with the generator refining its latent input using a temporal difference learning rule:
\begin{equation}
\mathbf{z}_{t+1} = \mathbf{z}_t + \alpha \cdot \delta_t \cdot \gamma^t \cdot \mathcal{N}(0, I),
\end{equation}
where $\alpha$ is a learning rate, $\gamma$ is a discount factor, and $\delta_t$ is the TD error computed by comparing step-wise and cumulative rewards.

During each step, the surrogate model $S$ evaluates the generated sample. For differentiable surrogates (e.g., neural networks), predictions can be directly used for gradient-based optimization. For non-differentiable surrogates (e.g., random forests), the predictions are averaged across temporal segments to obtain a class probability vector, from which we derive a binary label prediction:
\begin{equation}
\hat{\mathbf{y}} = 1 \left[ S(\tilde{\mathbf{x}}) > 0.5 \right]
\end{equation}
A step-wise reward is calculated by comparing $\hat{\mathbf{y}}$ to a randomly sampled or fixed target label $\mathbf{y}_{\text{target}}$, which measures how well the adversarial input can alter the model's decision. This reward guides the generator to produce inputs that are increasingly effective in inducing classification error.

At the end of each episode, the final latent input is used to compute a binary cross-entropy loss between surrogate predictions and pseudo-target labels:
\begin{equation}
\mathcal{L}_G = \text{BCE}(S(\tilde{\mathbf{x}}), \mathbf{y}_{\text{target}})
\end{equation}
The generator is updated using Adam optimizer to minimize this loss. This adversarial training loop continues over multiple episodes, enabling the generator to learn how to fool the surrogate classifier progressively.

\subsection{ML and DL Models}
We employ various machine learning and deep learning methods for insurance fraud detection. In this paper, we use LSTM, XGBoost, LightBoost, KNN, and SVM as baseline models.The details of the principles of different models are as follows.

\textit{LSTM}:Long Short-Term Memory (LSTM) networks are a type of Recurrent Neural Network (RNN) designed to address the vanishing gradient problem. They are particularly effective for classification tasks involving sequential data. The LSTM cell updates its internal state using the forget gate \( f_t \) and the input gate \( i_t \), which are computed as:
\begin{equation}
f_t = \sigma(W_f \cdot [h_{t-1}, x_t] + b_f)
\end{equation}
\begin{equation}
i_t = \sigma(W_i \cdot [h_{t-1}, x_t] + b_i)
\end{equation}
where \( \sigma \) is the sigmoid function, \( W \) represents weight matrices, \( b \) denotes bias terms, \( h_{t-1} \) is the previous hidden state, and \( x_t \) is the current input. This mechanism allows LSTMs to selectively retain or discard information, enhancing their ability to learn long-term dependencies.

\textit{XGBoost}:
XGBoost (eXtreme Gradient Boosting) is a powerful and efficient machine learning algorithm used for classification and regression tasks. It operates by building an ensemble of decision trees sequentially, where each tree corrects the errors of its predecessor. The objective is to minimize the loss function, which is defined as:
\begin{equation}
L(\phi) = \sum_{i=1}^{n} l(\hat{y}_i, y_i) + \Omega(\phi)
\end{equation}
where \( l \) is the loss function measuring the difference between predicted values \( \hat{y}_i \) and actual values \( y_i \), and \( \Omega \) is a regularization term that penalizes the complexity of the model to prevent overfitting. XGBoost uses gradient boosting to optimize this objective function, making it highly effective for various classification problems.

\textit{LightBoost}:LightGBM, short for Light Gradient Boosting Machine, is an efficient gradient boosting framework that excels in classification tasks. It constructs decision trees leaf-wise, optimizing for the leaf that maximizes the decrease in loss. This method contrasts with level-wise growth, enhancing efficiency. LightGBM also utilizes a histogram-based algorithm for faster training. The objective function is given by:
\begin{equation}
L(\phi) = \sum_{i=1}^{n} l(\hat{y}_i, y_i) + \Omega(\phi)
\end{equation}
Where \( l \) calculates the discrepancy between predictions \( \hat{y}_i \) and true values \( y_i \), and \( \Omega \) regularizes the model to avoid overfitting. This setup makes LightGBM both accurate and computationally efficient.

\textit{KNN}:K-Nearest Neighbors (KNN) is a simple yet effective algorithm used for classification tasks. It operates on the principle of majority voting among the k nearest neighbors in the feature space. Given a new data point, KNN identifies the k most similar instances from the training dataset based on a distance metric, typically Euclidean distance. The classification is determined by the majority class among these neighbors. The decision boundary is defined by the equation:
\begin{equation}
d(x, x_i) = \sqrt{\sum_{j=1}^{n} (x_j - x_{ij})^2}
\end{equation}
where \(d(x, x_i)\) represents the distance between the new data point \(x\) and a training instance \(x_i\), and \(n\) is the number of features. KNN is easy to implement and interpret but can be computationally intensive for large datasets due to its reliance on distance calculations.

\textit{SVM}:Support Vector Machine (SVM) is a classifier that operates on the principle of finding the optimal hyperplane that maximally separates different classes in the feature space. The goal is to maximize the margin, which is the distance between the hyperplane and the nearest data points from any class, known as support vectors. The decision function for SVM is given by:
\begin{equation}
f(x) = \text{sign} \left( \sum_{i=1}^{n} y_i \alpha_i \langle x_i, x \rangle + b \right)
\end{equation}
where \( \langle x_i, x \rangle \) represents the dot product between the support vectors \( x_i \) and the input vector \( x \), \( y_i \) are the class labels, \( \alpha_i \) are the Lagrange multipliers, and \( b \) is the bias term. SVM can handle high-dimensional data and is effective in cases where the number of dimensions exceeds the number of samples.

\section{Experiment and Results}
The insurance fraud detection systems are assessed using accuracy, F1 score, and Attack Success Rate (ASR). The formulas are as follows:
\begin{equation}
\text{Accuracy} = \frac{TP + TN}{TP + FP + TN + FN},
\end{equation}
\begin{equation}
F1 = \frac{2TP}{2TP + FP + FN},
\end{equation}
\begin{equation}
ASR = \frac{\text{\# batches deceiving stability}}{\text{\# batches sent}}.
\end{equation}

\subsection{Dataset}
\begin{figure}[h]
    \centering
    \includegraphics[width=0.9\textwidth]{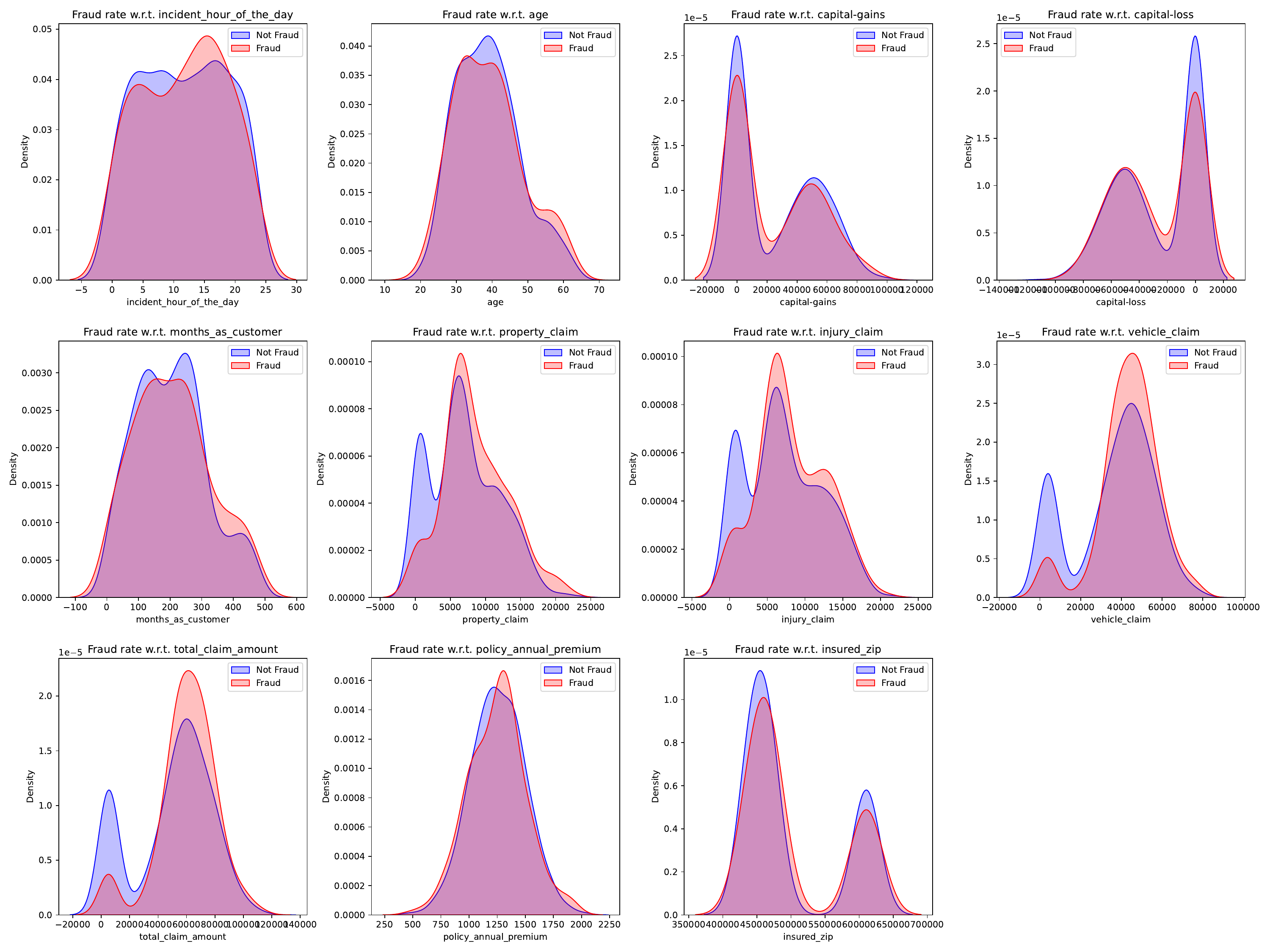} 
    \caption{Distribution of characteristics of different categories}
    \label{fig:cdf}
\end{figure}

This research utilizes a publicly available insurance dataset, which contains 1,000 samples and 38 features. The distribution of numerical features across different classes is shown in the figure \ref{fig:cdf}. Additionally, we divided the dataset into training (75\%), validation (5\%), and test (20\%) subsets. We performed normalization during preprocessing to effectively prepare the data for the prediction models.



\subsection{Setting and Baselines}
To ensure practicality and efficiency in insurance fraud detection, the deep learning model was engineered to be streamlined, minimizing computational demands while maximizing effectiveness. This design philosophy aligns with the goal of creating a robust yet resource-efficient system. The insurance fraud detection model employs a one-layer Bi-directional LSTM architecture with 220 neurons to capture temporal dependencies in both forward and backward directions within the claim sequence. To prevent overfitting, a dropout layer with a 0.5 dropout rate was introduced during training. Following the dropout layer, the LSTM layer's output passes through a Linear layer with 440 neurons, activating an element-wise sigmoid function. The deliberate choice of LSTMs aims to capture potential causal relationships between claim data points. For model optimization, the Binary Cross-Entropy loss function, a standard metric for binary classification tasks, was used. The training process utilized the Adam optimizer with a learning rate of $1 \times 10^{-3}$ for efficient gradient descent. Training iterations were structured into 10 epochs to balance duration and performance.

The parameter settings for the GAN are as follows: To enhance gradient propagation and avoid the dying ReLU issue, Leaky ReLU activations are adopted\cite{lim2024future}\cite{rahman2024syn}. The generator consists of five fully connected layers with 128, 256, 512, 64, and 12 units, which facilitate stable training and effective feature transformation. To thoroughly assess the efficacy of the method introduced in this paper, a selection of prevalent methodologies were chosen as baselines. These include FGSM\cite{goodfellow2014explaining}, BIM\cite{kurakin2016adversarial}, PGD\cite{madry2017towards}, and Random Noise. The methods of the baselines are described in detail as follows:

\begin{itemize}
    \item \textit{Fast Gradient Sign Method (FGSM)}: Adversarial examples are generated using the sign of the loss gradient. Known for its high computational efficiency, FGSM is commonly used to assess model robustness.
    \item \textit{Basic Iterative Method (BIM)}: FGSM is extended by applying multiple small perturbations, thereby increasing the attack's strength. This method emphasizes the cumulative effect on model robustness.
    \item \textit{Projected Gradient Descent (PGD)}: BIM is enhanced by incorporating a projection step to confine perturbations within a specified constraint set, allowing for more powerful attacks and a more thorough evaluation of robustness.
    \item \textit{Random Noise Attack}: Noise is introduced to create adversarial examples that compromise models. Perturbations are drawn from a normal distribution, guided by the user-defined epsilon ($\epsilon$) parameter, which controls the attack's intensity and the extent of perturbation. A larger $\epsilon$ leads to more substantial changes, potentially facilitating successful deception. In contrast, a smaller $\epsilon$ produces more subtle perturbations that are harder to detect but can still be effective. After perturbation, samples are classified. If the accuracy decreases below the original, the perturbed sample replaces the original in the adversarial set. This iterative process continues until a successful adversarial instance is identified.
\end{itemize}

\subsection{Results and SHAP Value}
\begin{figure}
    \centering
    \includegraphics[width=0.9\linewidth]{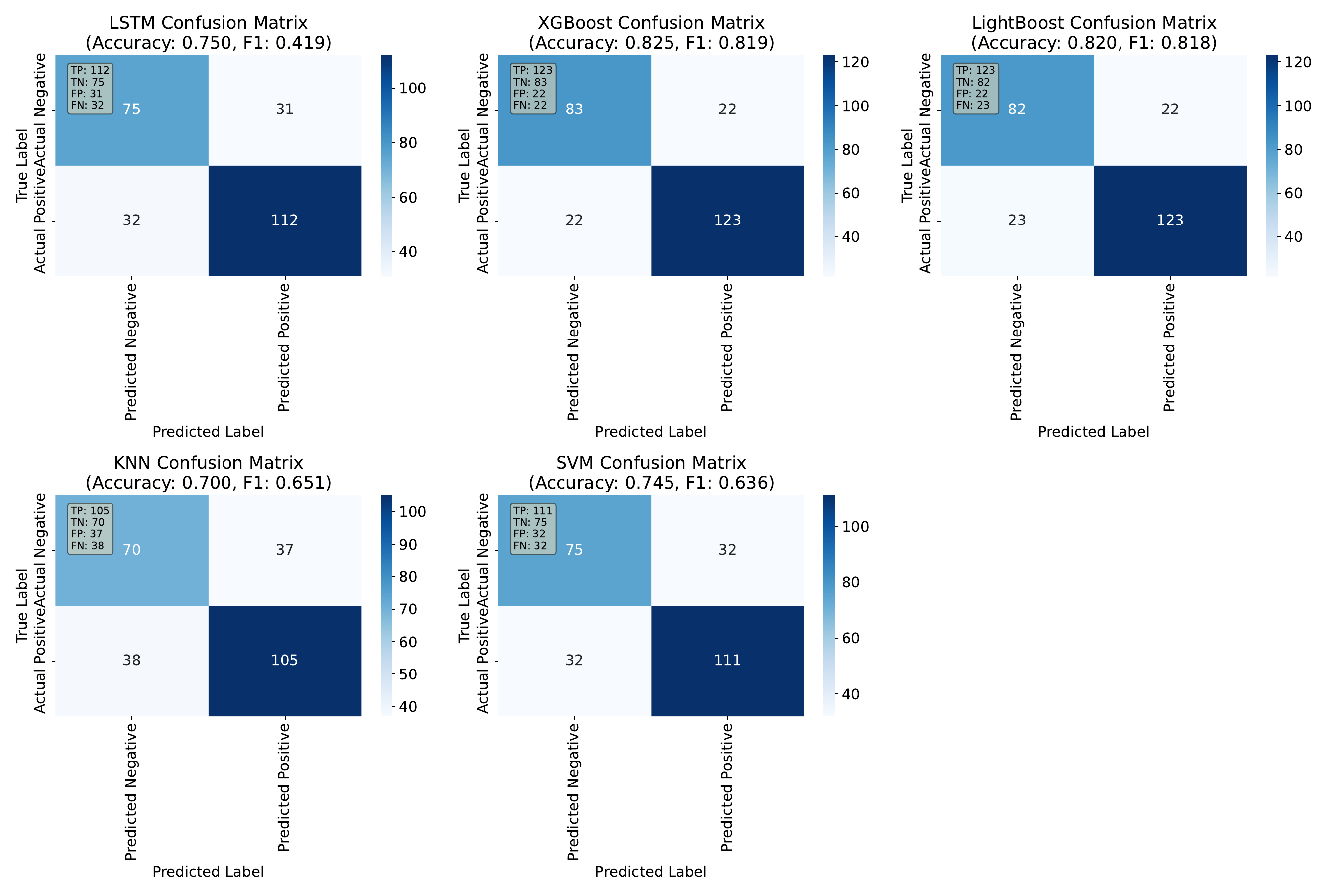}
    \caption{Confusion matrices of Different Models}
    \label{fig:matrix}
\end{figure}

The result can be seen in table \ref{tab:res_class}.From the results in the table, it can be observed that the XGBoost model performed the best, with an Accuracy of 0.825 and an F1 Score of 0.819. The LightBoost model ranked second, achieving an Accuracy of 0.820 and an F1 Score of 0.819. It is important to note that although the LSTM model reached an Accuracy of 0.750, its F1 Score was only 0.419. This discrepancy may be due to the LSTM model's strength in capturing temporal relationships, which generally results in its predictive performance being lower than that of machine learning models.The confusion matrices of different models are shown in the figure \ref{fig:matrix}.
\begin{table}[h]
\centering
\caption{Performance Metrics of Different Models}
\begin{tabular}{@{}l|cc@{}}
\toprule
Model & Accuracy & F1 Score \\ \midrule
LSTM & 0.750 & 0.419 \\
XGBoost & 0.825 & 0.819 \\
LightBoost & 0.820 & 0.818 \\
KNN & 0.700 & 0.651 \\
SVM & 0.745 & 0.636 \\\bottomrule
\end{tabular}
\label{tab:res_class}
\end{table}

\begin{figure}[h]
    \centering
    \begin{subfigure}[b]{0.45\textwidth}
        \includegraphics[width=\textwidth]{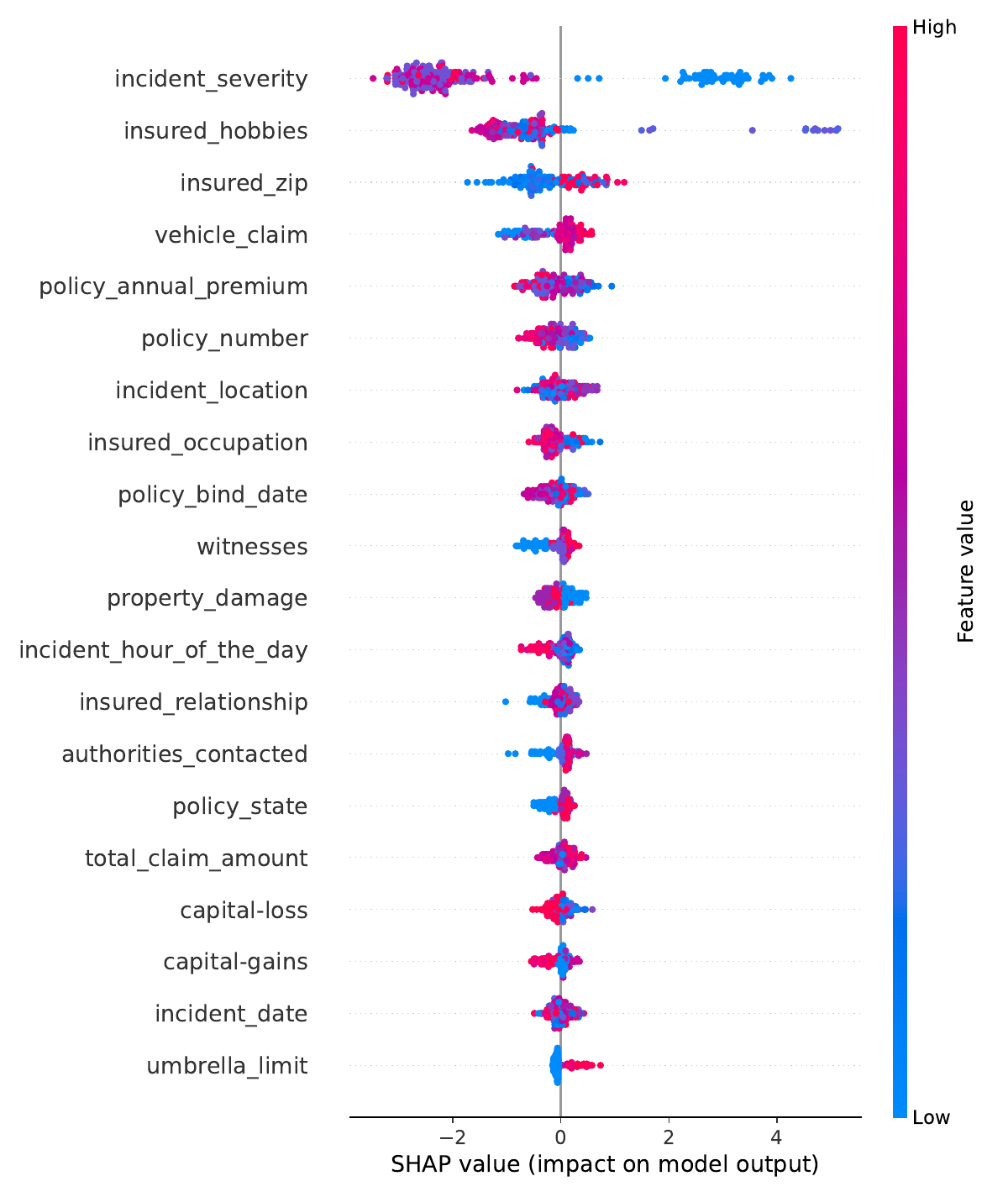}
        \caption{LSTM}
        \label{fig:shap_xgb}
    \end{subfigure}
    \hfill 
    \begin{subfigure}[b]{0.45\textwidth}
        \includegraphics[width=\textwidth]{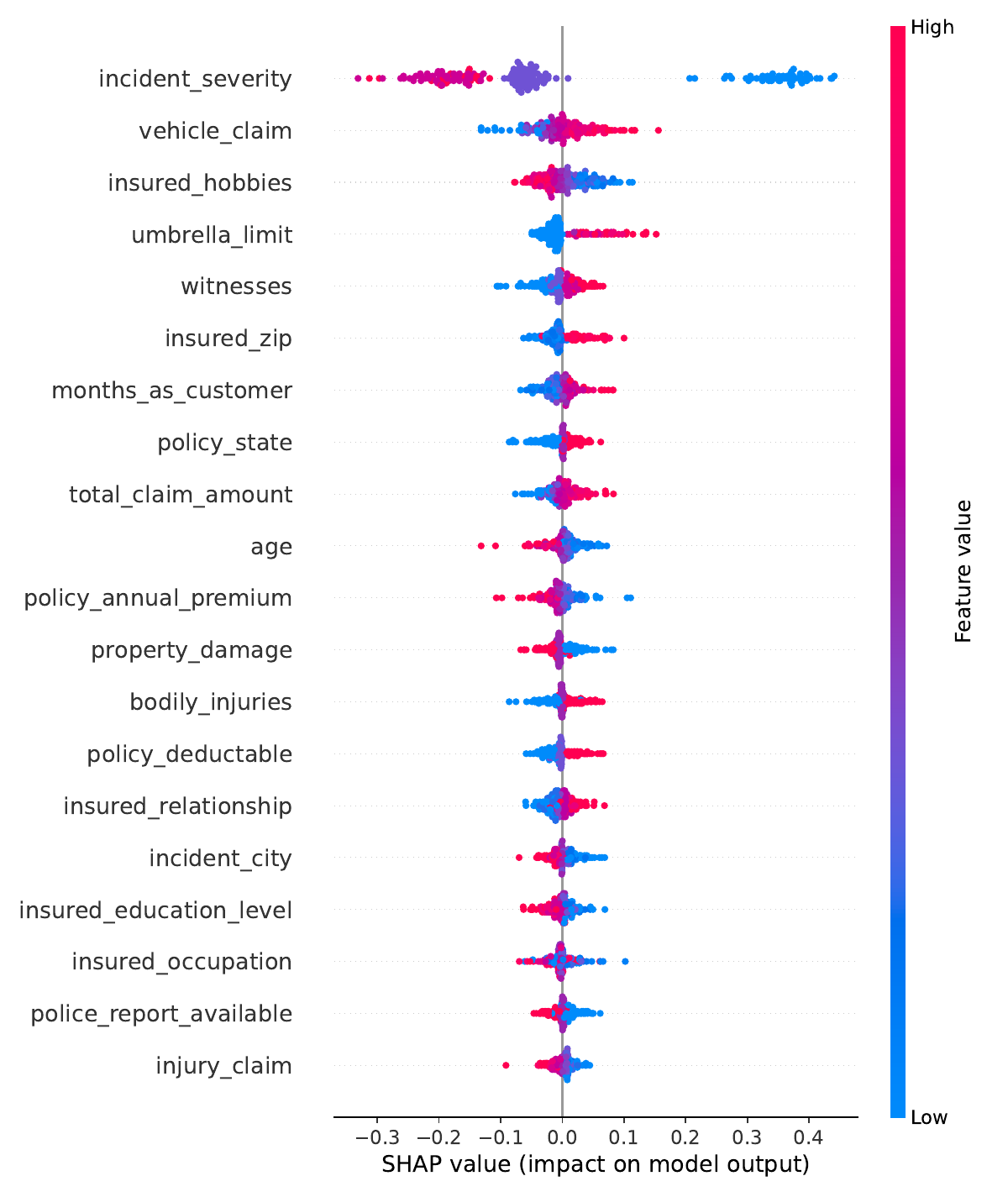}
        \caption{XGBoost}
        \label{fig:shap_lstm}
    \end{subfigure}
    \caption{SHAP values for LSTM and XGBoost}
    \label{fig:shap}
\end{figure}

To further analyze the results of insurance fraud detection, we employed Explainable Artificial Intelligence (XAI) techniques to explore feature importance. Specifically, we used the SHAP (Shapley Additive exPlanations) values method. The SHAP method can quantify the contribution of each feature to the model's predictions and provide interpretability of the decision-making process. We used SHAP Gradient Explainer to interpret the LSTM model and SHAP Tree Explainer for both XGBoost and LSTM models, with the results shown in figure \ref{fig:shap_xgb} and figure \ref{fig:shap_lstm}.
From the figure \ref{fig:shap}, we can observe that LSTM and XGBoost assign different importances to various features. Specifically, both LSTM and XGBoost have learned that the most important feature is incident\_severity. Beyond this, LSTM considers the next three most important features to be insured\_hobbies, insured\_zip, and vehicle\_claim, while XGBoost identifies the top three features as vehicle\_claim, insured\_hobbies, and umbrella\_limit.The varying degrees of attention to features by different machine learning models affect the final classification results.

\subsection{Attack Analysis}
\begin{figure}[h]
    \centering
    \includegraphics[width=0.75\textwidth]{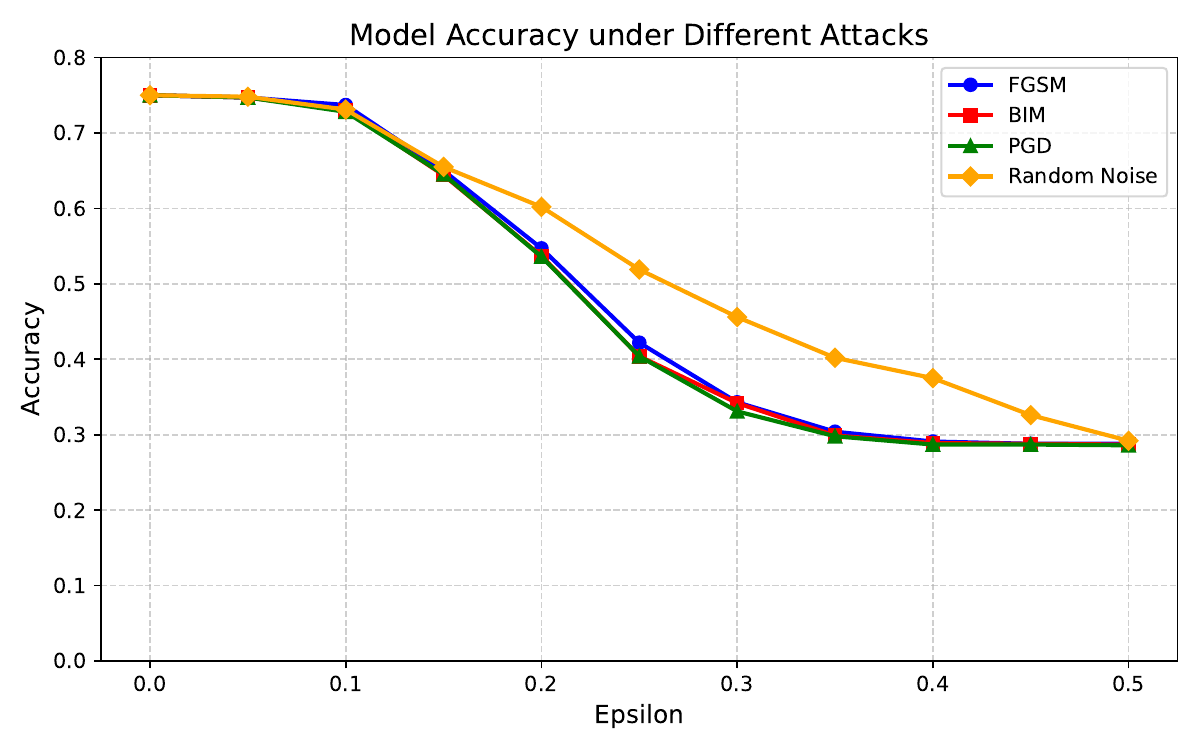} 
    \caption{Model accuracy vs epsilon for LSTM.}
    \label{fig:attack_lstm}
\end{figure}

\begin{figure}[h]
    \centering
    \includegraphics[width=0.75\textwidth]{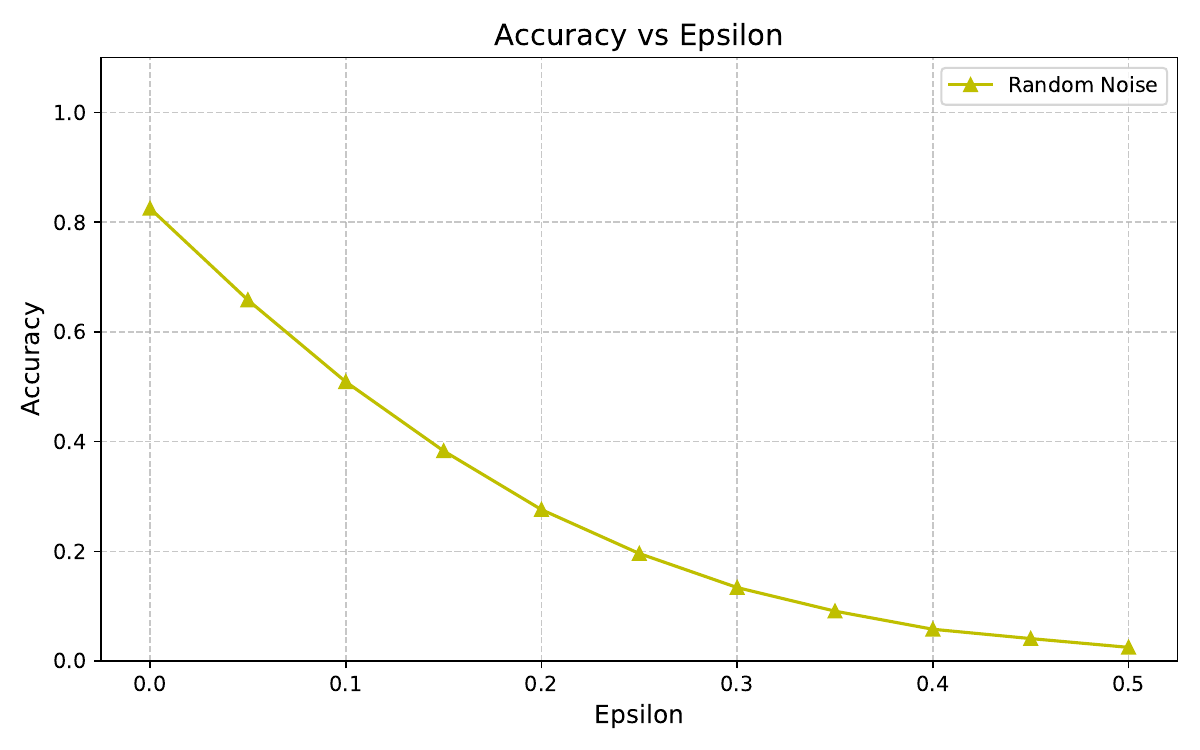} 
    \caption{Model accuracy vs epsilon for XGBoost.}
    \label{fig:attack_xgb}
\end{figure}

In this section, we explored the effectiveness of white-box attacks. We used the Adversarial Robustness Toolbox (ART) library\cite{nicolae2018adversarial} to probe the baseline system and assess the vulnerability of the model without any defense measures. For classical machine learning models with non-differentiable architectures, such as decision trees or ensemble methods, applying white-box adversarial attacks (e.g., FGSM, BIM, and PGD) is not straightforward. Obtaining gradients is a key challenge. Unlike deep learning models that can conveniently provide gradients, classical machine learning models often cannot do so. This makes gradient-based attacks either difficult to implement or fraught with challenges. This difficulty also reflects the fundamental differences in architecture and methods between classical machine learning and deep learning. Traditional machine learning techniques often rely on discrete decision-making and non-linear transformations, which complicate the calculation and propagation of gradients. Moreover, existing implementations of attack libraries do not support machine learning classifiers. Therefore, we did not apply these three types of attacks to traditional machine learning models. Instead, we employed a random noise attack tailored for XGBoost to explore potential security vulnerabilities and assess robustness. The attacks were conducted with varying epsilon values, which represent the strength of each attack and the extent of perturbation introduced. We investigated epsilon values ranging from 0.05 to 0.50. The outcomes of these attacks across different models are visually depicted in Figure \ref{fig:attack_lstm} and Figure \ref{fig:attack_xgb}. Compared to the LSTM model, the XGBoost model is more susceptible to the same random noise attack. Additionally, it is noteworthy that the FGSM, BIM, and PGD methods exhibit nearly identical performance, surpassing that of random noise. Moreover, increasing the epsilon value beyond 0.5 does not provide any significant advantage.

The method proposed in this paper employs a GAN approach optimized through reinforcement learning. The objective is to generate data that can be detected as non-fraudulent by the model without actual access to the model architecture. The results show an attack success rate (ASR) of 0.99 for both LSTM and XGBoost models, highlighting the vulnerability of current insurance fraud detection models. Table \ref{tab:accuracy} illustrates the model accuracy under different attack methods, where the baseline method requires access to both the model and data, which may pose significant challenges in the real world. In contrast, the method proposed in this paper can effectively launch attacks by accessing only the model output. In terms of attack success rate, the method proposed in this paper outperforms other methods.
\begin{table}[h]
\centering
\begin{tabular}{lcccccc}
\toprule
Model & \multicolumn{6}{c}{Accuracy} \\
\cmidrule(lr){2-7}
 & Baseline & Our & FGSM & BIM & PGD & Random noise \\
\midrule
LSTM & 0.750 & 0.01 & 0.295 & 0.294 & 0.297 & 0.295 \\
XGBoost & 0.814 & 0.01 & -- & -- & -- & 0.07 \\
\bottomrule
\end{tabular}
\caption{Comparison of Model Performance Under Adversarial Attacks($\epsilon$=0.5).}
\label{tab:accuracy}
\end{table}

\section{Conclusion}
This paper presents a novel GAN-based adversarial attack method specifically designed to target insurance fraud detection systems. Leveraging the generative power of GANs, the proposed approach is optimized using reinforcement learning to produce highly deceptive fraudulent cases capable of bypassing state-of-the-art detection models. Experiments demonstrate an exceptionally high attack success rate (ASR) of 99\% on both LSTM and XGBoost models, revealing a critical vulnerability in current fraud detection systems. Notably, these attacks are successful even under constrained scenarios with limited access to training data or model architecture, which raises serious concerns about the robustness of widely deployed models in real-world insurance applications.The results highlight a pressing need to reassess the security assumptions underlying existing fraud detection frameworks. As adversarial techniques become increasingly sophisticated, traditional machine learning and deep learning models may fall short in providing adequate protection. The demonstrated effectiveness of this attack poses a serious threat to the reliability and stability of insurance infrastructures, potentially leading to significant financial losses and erosion of trust in automated fraud screening processes.Future work should prioritize the development of robust adversarial defense mechanisms, such as adversarial training, detection of anomalous input patterns, and uncertainty-aware models. Strengthening these systems is essential to safeguard the integrity of insurance fraud detection and maintain industry resilience.

\bibliographystyle{plain}
\bibliography{reference}

\end{document}